\documentclass[a4paper,fleqn,usenatbib]{mnras}

\usepackage[T1]{fontenc}
\usepackage{ae,aecompl}

\usepackage{graphicx}	
\usepackage{amsmath}	
\usepackage{amssymb}	

\title[Nucleosynthesis in the $R_h = ct$ cosmology]{Primordial Nucleosynthesis in the $R_h = ct$ cosmology: \\
Pouring cold water on the Simmering Universe}

\author[G. F. Lewis et al.]{
Geraint F. Lewis\thanks{E-mail: geraint.lewis@sydney.edu.au (GFL)},
Luke A. Barnes,
and Rajesh Kaushik
\\
Sydney Institute for Astronomy, School of Physics, A28, The University of Sydney, NSW 2006, Australia 
}

\date{Accepted XXX. Received YYY; in original form ZZZ}

\pubyear{2016}

\begin{document}
\label{firstpage}
\pagerange{\pageref{firstpage}--\pageref{lastpage}}
\maketitle

\begin{abstract}
Primordial nucleosynthesis is rightly hailed as one of the great successes of the standard cosmological model. Here we consider the initial forging of elements in the recently proposed $R_h = ct$ universe, a cosmology that demands linear evolution of the scale factor. 
Such a universe cools extremely slowly compared to standard cosmologies, considerably depleting the available neutrons during nucleosynthesis; 
this has significant implications for the resultant primordial abundances of elements,
predicting a minuscule quantity of helium which is profoundly at odds with observations. 
The production of helium can be enhanced in such a ``simmering universe'' by boosting the baryon to photon ratio, although more than an order of magnitude 
increase is required to bring the helium mass fraction into accordance with observations.
However, in this scenario, the prolonged period of nucleosynthesis results of the efficient cooking of lighter  into heavier elements, 
impacting the resultant abundances of all elements so that,  other than hydrogen and helium, there are virtually no light elements  present in the universe.
Without the addition of substantial new physics in the early universe, it is difficult to see how the $R_h = ct$ universe can be considered a viable  cosmological model.
\end{abstract}

\begin{keywords}
cosmology: theory
\end{keywords}



\section{Introduction}\label{introduction}
 One of the earliest successes of the standard cosmological model arose from the realisation that the early universe was hot and dense and that within these conditions  some of the lighter elements can be forged from the initial soup of baryons, leptons and photons \citep{1948PhRv...74.1198A,1964Natur.203.1108H}
However, the period for element formation was relatively short, with the rapidly cooling universe shutting off nucleosynthesis within the first 20 minutes after the Big Bang \citep{1966ApJ...146..542P,1967ApJ...148....3W,1968ApJ...151L.103W}. The short cooking window predicts a primordial universe dominated by hydrogen, with a helium mass fraction of 25\% and trace amounts of other elements, in amazing agreement with the observed abundances \citep[see][for a comprehensive review of the physics of the early universe]{1990eaun.book.....K}.

Even though it has proved to be very successful, the $\Lambda$CDM ($\Lambda$ Cold Dark Matter) paradigm is regularly challenged. One of the more recent newcomers is the so-called $R_h = ct$ universe \citep{2009IJMPD..18.1889M,2012MNRAS.419.2579M}, founded on the claim that the "Hubble Sphere" \citep{1991ApJ...383...60H} is an unrecognised horizon in the universe \citep{2007MNRAS.382.1917M}, although this been shown to be incorrect~\citep{2010MNRAS.404.1633V,2012MNRAS.423L..26L}. This model also encompassed a numerological coincidence, namely that the age of the universe, $t_o$, and the present day Hubble constant, $H_o$, are related via $H_o t_o \sim 1$.  Within $\Lambda$CDM, such a coincidence occurs only for a short period of its infinite lifetime. However, if we suppose this coincidence is telling us something deeper about the universe, and demand that $H t = 1$ precisely at all points in the universe's history, we must demand that scale factor must evolve as $a(t) \propto t$, a very different cosmic history to $\Lambda$CDM. 
Before continuing, however, it is worth noting that this numerological relationship is only approximately true given our recent determination of cosmological parameters, with $H_o t_o = 1.05 \pm 0.2$, providing a significant challenge to the underlying motivation of the $R_h = ct$ cosmological model \citep{2015mgm..conf.1567V}. 

This has not halted the continued appearance of this cosmology and it has been claimed that such a linearly expanding universe provides a better description of the observed universe \citep[e.g.][]{2012AJ....144..110M,2013A&A...553A..76M,2014MNRAS.442..382M,2015AJ....150..119M,2016MNRAS.456.3422M}, a claim that has met with resistance 
\citep[e.g. ][]{2012MNRAS.425.1664B,2014MNRAS.442..382M,2013MNRAS.432.2324L,2015PhRvD..91j3516S}, and possesses significant underlying physical constraints, demanding interacting component to maintain a `zero active mass' condition to achieve linear expansion [e.g. \citet{2016FrPhy..11..557M} and \citet{2016arXiv160201435M}; although see \citet{2016arXiv160107890K}], it continues to receive significant attention from some. 

In this paper, we consider the implications of demanding the linear expansion of the $R_h = ct$ universe on the initial stages of the universe, in particular during the epoch of nucleosynthesis. In Section~\ref{nucleosynthesis} we present a discussion of standard primordial nucleosynthesis as well as detailing our investigation of nucleosynthesis in the $R_h = ct$, with a theoretical exploration presented in Section~\ref{melia} and a numerical solution to the nucleosynthesis equations presented in Section~\ref{numerical}. In Section~\ref{powerlaw} we review the recent history of power-law cosmologies, finding that objections presented in favour of and against the $R_h = ct$ cosmology have appeared in previous literature. We present our conclusions in Section~\ref{conclusions}.

\section{Primordial Nucleosynthesis}\label{nucleosynthesis}

\subsection{Numerical Calculations}\label{numerical}
In the following we will present a combination of analytic and numerical arguments to examine the impact of demanding a linear expansion of the universe during the period of primordial nucleosynthesis. 
For this we employ a modified version of {\tt AlterBBN}, which simulates Big Bang Nucleosynthesis by numerically integrating the yield equations for the various nuclear interactions occurring in the early universe  \citep{2012CoPhC.183.1822A}. This code can consider a number of modifications to the physics of the early stages of the universe, but for our purposes we make two simple modifications. To enforce a linear evolution of the scale factor, the Hubble constant is equal to 
\begin{equation}
H = \frac{1}{t}
\label{hubble}
\end{equation}
where $t$ is the age of the universe at the epoch under consideration. Given this, the temperature of the at this time is given by
\begin{equation}
T = \frac{T_o}{ H_o t }
\label{temperature}
\end{equation}
where $H_o$ is the present day value of the Hubble constant, and $T_o$ is the present day temperature of the cosmic microwave background, taken to be $T_o = 2.725 K$;
note that in the following we have not considered the effective number of particle species which modifies the temperature profile in the early universe, although we find that this has no impact on the conclusions presented in this paper.  As well as our modified cosmological calculations with {\tt AlterBBN}, we also undertook a standard cosmological run to provide a fiducial comparison.

\subsection{Standard Big Bang Nucleosynthesis}\label{standard}
As noted, Big Bang nucleosynthesis is one of the great successes of cosmology of the last century, and the physical details are presented in numerous sources \citep[e.g.][]{1990eaun.book.....K,2006AIPC..843..111P} as well as being the staple of undergraduate courses on cosmology. Here we will briefly review the key points\footnote{For the interested reader, we recommend following the detailed steps laid out in \citet{2005pfc..book.....M}}.

\begin{figure}
		\includegraphics[width=0.95\columnwidth]{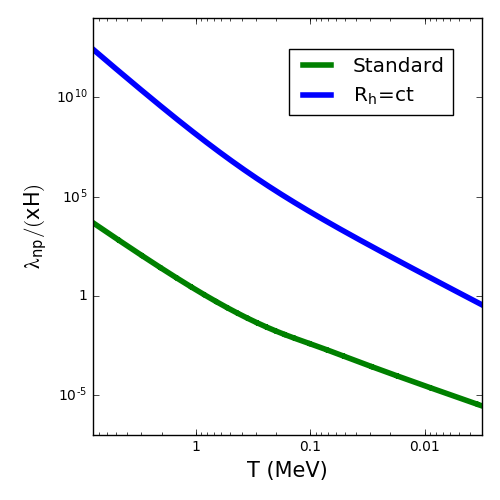}
    \caption{The neutron-to-proton conversion rate, relative to the expansion of the universe (see Eqn.~\ref{dyneqn}) for the standard cosmological model (green line) and the $R_h = ct$ cosmology (blue line). Note that this has a value of roughly unity at 1 MeV, diminishing quickly as the universe cools to lower energy, ensuring a freeze-out of the neutron to proton ration. At 1 MeV in the $R_h = ct$ cosmology, this has a corresponding value eight orders of magnitude larger, not decline to unity until temperatures are well below 0.01 MeV. This ensures that there is no freeze-out of the proton to neutron ratio before nucleosynthesis in this cosmological model.}
    \label{freezeoutplot}
\end{figure}

Within the standard cosmological model, primordial nucleosynthesis occurs in the seconds to minutes after the Big Bang.  
Given that the diproton is unbound, the build up of nuclei is dependent on the availability of free neutrons, which allow the formation of deuterium ($^\text{2}$H or D) and thereafter the heavier elements. To examine this, we define
\begin{equation}
X_n \equiv \frac{ n_n }{ n_n + n_p }
\label{numberratio}
\end{equation}
The expression for the  dynamical equation for the evolution of $X_n$ is given by
\begin{equation}
\frac{d X_n}{dx} = \frac{\lambda_{np}}{x H} \left( (1-X_n)e^{-x} - X_n \right)
\label{dyneqn}
\end{equation}
where $x = {\cal Q}/T$, ${\cal Q} = m_n - m_p$, and $m_n$ and $m_p$ are the masses of the neutron and proton respectively, and $T$ is the temperature of the universe (assuming $c=k=1$). Furthermore,  $\lambda_{np}$ is the neutron to proton conversion rate and is given by 
\begin{equation}
\lambda_{np} = \frac{255}{\tau_n x^5} \left( x^2 + 6x + 12\right)
\label{pnrate}
\end{equation}
where $\tau_n$ is the neutron lifetime. If $\lambda_{np}$ is fast compared to the Hubble expansion, the thermonuclear ratio of neutrons to protons, given by 
\begin{equation}
\frac{n_n}{n_p} = \left( \frac{ m_n }{ m_p } \right)^\frac{3}{2} e^{-\frac{\cal{Q}}{T}}
\label{pnratio}
\end{equation}
is maintained, meaning that, as the temperature drops, the relative availability of neutrons rapidly diminishes. However, 
in the standard cosmological model, this is not the case (c.f. Figure~\ref{freezeoutplot} and discussion in Section~\ref{melia}); here
the expansion is dominated by the presence of the radiative components, with the expansion going as  $a(t) \propto t^\frac{1}{2}$ and as it cools below $\sim \text{0.7 MeV}$, the rates of $p \leftrightarrow n$ become slow compared to expansion and the ratio of neutrons to protons is "frozen-out" with a ratio of;
\begin{equation}
\frac{n_n}{n_p} \sim \frac{1}{6}
\label{frozen}
\end{equation}
At this period, the universe is still too hot for significant nucleosynthesis, entering what is known as the "deuterium bottleneck", which requires a temperature below $\sim\text{0.07 MeV}$ for $^\text{2}\text{H}$ formation to dominate over destruction, and a proportion of neutrons undergo weak decay through $n \rightarrow p + e^- + \bar{\nu}_e$. A proportion of neutrons remain to be bound into deuterium and then helium such that the final mass fraction of $^\text{4}\text{He}$ is $\sim 25\%$.

\subsection{Neutron-Proton Freeze-out in the $R_h = ct$ Universe}\label{melia}
It is straightforward to consider the above analysis in light of the $R_h = ct$ cosmological model. The astounding thing about this relationship is that the age of the universe at the epoch of nucleosynthesis  (T$\sim10^9 - 10^7$ K) is about $10^9$ seconds, corresponds to about 30 years! Clearly this age will have an impact on the resultant nucleosynthesis, begin with the freeze-out (or, as we will see, lack of freeze-out) in the $R_h = ct$ universe.

We can consider calculating the the appropriate rates for the $R_h = ct$ universe in the same fashion as for the standard cosmological model (Section~\ref{standard}). 
Firstly, the relevant ratio of the neutron to proton conversion rate to the Hubble expansion (Eqn.~\ref{pnrate}) is given by
\begin{equation}
\frac{\lambda_{np}}{x H} = \frac{255 H_o}{\tau_n x^6}\frac{T}{T_o} \left(   x^2 + 6x + 12  \right)
\label{stuff}
\end{equation} 
We plot this in Figure~\ref{freezeoutplot}, with the blue line representing the $R_h = ct$ universe, whereas the green line is the standard cosmological model. At a temperature of $\sim$1 MeV in the standard cosmological model, this term is roughly unity, dropping rapidly as the universe continues to cool, and ensuring a freeze-out of the neutron to proton ration before the onset of nucleosynthesis. However, in the $R_h = ct$ cosmology, the corresponding term is roughly eight orders of magnitude higher at $\sim$1 MeV, not dropping to unity until the temperature of the universe is well below $\sim$0.01 MeV. The direct consequence of this is that the neutron to proton ratio maintains thermodynamic equilibrium into the epoch of nucleosynthesis, leading to a significant depletion in available neutrons for the build up of nuclei.

\begin{figure}
		\includegraphics[width=0.95\columnwidth]{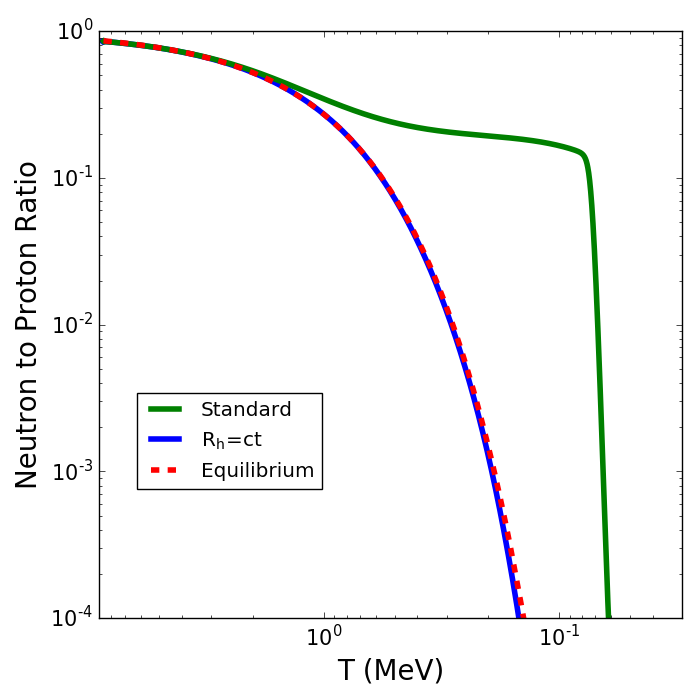}
    \caption{The neutron to proton ratio within the standard cosmological model (green line) and the $R_h = ct$ universe (blue line), while the red dashed line corresponds to the equilibrium ratio given by Equation~\ref{dyneqn}. At temperatures above $\sim$1 MeV, the neutron to proton ratio in both cosmologies is in equilibrium, but as the standard universe cools, the neutron to proton ratio freezes out, departing from equilibrium, and then declines due to neutron decay. The precipitous decline at T$\sim$ 0.05 MeV is the onset of nucleosynthesis where all free neutrons are bound up into deuterium and helium. Conversely, in the $R_h = ct$ universe, the slow cooling and steady expansion ensures that the neutron to proton ratio does not depart from its equilibrium value, significantly depleting the universe of available neutrons.}
    \label{freezeoutplot2}
\end{figure}

To further emphasise this point, in Figure~\ref{freezeoutplot2} we plot the neutron to proton ratio in both the standard cosmological model, and the $R_h = ct$ universe, as well as the theoretical equilibrium value (Eqn.~\ref{pnratio}), demonstrating the dramatic depletion of neutrons in the linearly expanding universe. As net production of deuterium ($^\text{2}$H) is not significant until the temperature has decreased to $\sim$0.2 MeV, the expected neutron to proton ratio at that time will be $\sim 2 \times 10^{-3}$. Assuming that all neutrons get finally bound into $^\text{4}$He, the resulting abundance of primordial helium will be $\sim 4 \times 10^{-3}$, well below the observed value.

\subsection{Primordial Abundances in the $R_h = ct$ Universe}\label{abundances}
We have seen that, at the onset of nucleosynthesis, the neutron abundance in the $R_h = ct$ cosmology has been significantly depleted due to the maintaining of equilibrium as the universe cools. Here we focus on the impact of this on the resultant elemental abundances after nucleosynthesis. For this, we numerically integrate the yield equations in {\tt AlterBBN}, remembering that this takes into account the physics leading to the lack of a freeze-out of the neutron to proton ratio discussed in Section~\ref{melia}\footnote{ For these calculations, we removed an approximation from the original code which cuts off the proton to neutron interaction below a limiting temperature  to expedite numerical calculations (Arbey, private communication)}.

\begin{figure}
		\includegraphics[width=0.99\columnwidth]{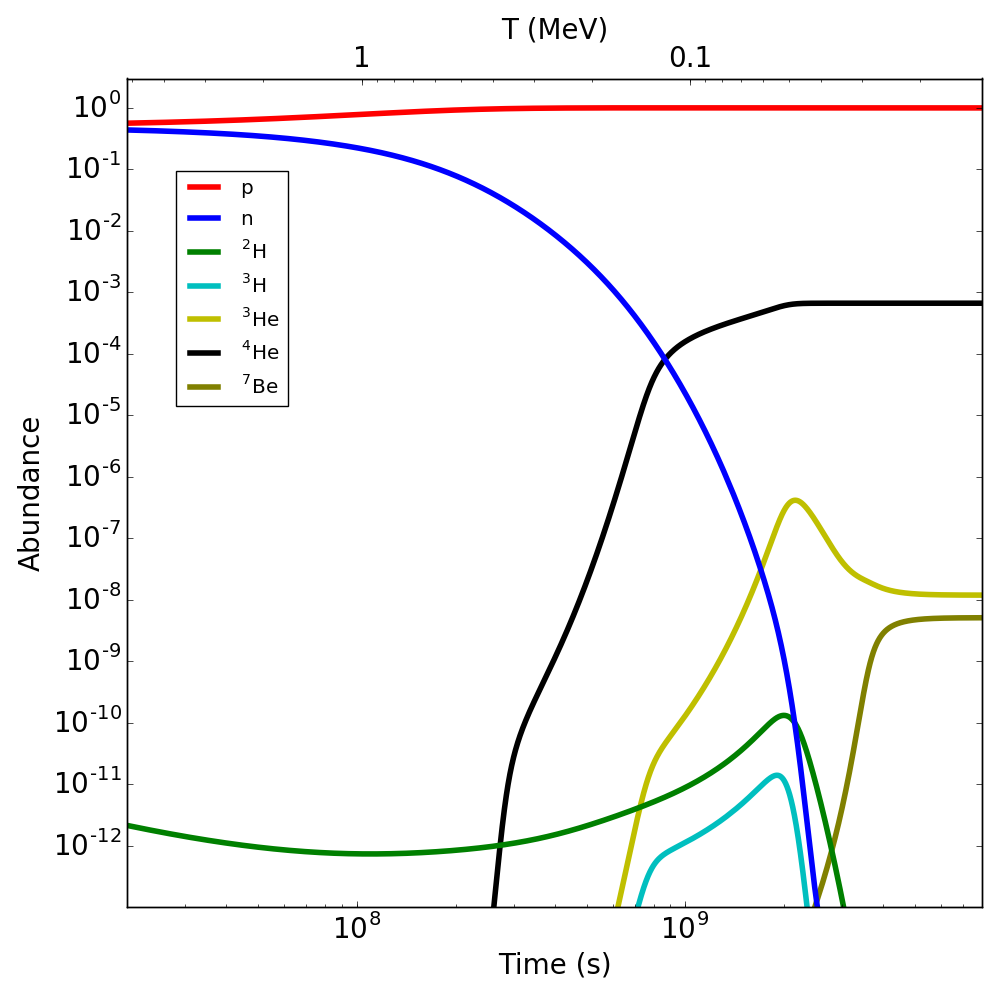}
    \caption{The cosmic abundance of elements verses time in the $R_h = ct$ cosmology; note the scale on the lower x-axis which corresponds to multiple years in the age of the universe. The upper x-axis presents the corresponding temperature of the universe (in MeV).}
    \label{abundance_vs_time}
\end{figure}

Figure~\ref{abundance_vs_time} presents the results of this numerical integration, presenting the cosmic abundance of elements as a function of time (the upper x-axis presents the corresponding temperature in MeV). As expected, the time-scale for nucleosynthesis in the $R_h = ct$ cosmology is over a period of years and decades, rather than the seconds and minutes seen in standard cosmological models. 
As expected from the analysis presented in Section~\ref{melia}, the universe is in equilibrium until about $\sim 10^8$s, after which the production of ${\rm ^2H}$ begins to exceed its destruction rate, and nucleosynthesis is underway in ernest by around $\sim 2\times10^8$s. The lack of a freeze-out in the neutron to proton ratio is apparent, with a relative abundance of 
$\sim 10^{-2} - 10^{-3}$ by this point. This deficit of neutrons feeds directly into the subsequent production of\ ${\rm ^2H}$ and then ${\rm ^4He}$, such that when nucleosynthesis is complete the resulting abundance of ${\rm ^4He}$ is $\sim 10^{-3}$. 
Examining the other elements reveals that, while trace amounts of $^\text{3}$He and $^\text{7}$Be are presents with abundances of $\sim10^{-8}$ and $\sim5\times10^{-9}$ respectively, values in themselves, quite different to the observed abundances, deuterium is completely absent, having been used up in subsequent nucleosynthesis. This is in conflict with the observed primordial deuterium abundance of (D/H)$_p$ = $(2.53 \pm 0.04) \times 10^{-5}$ \citep{2014ApJ...781...31C}. 
If the $R_h = ct$ cosmology is to be taken as a valid cosmological model, this result is extremely problematic.

\section{Power-law Cosmologies}\label{powerlaw}
We note that the $R_h = ct$ cosmology is not a new idea, but a resurgence of discussion of power-law cosmologies of the general form $a(t) \propto t^{\alpha}$, where $\alpha = 1$ gives the linear expansion of the $R_h = ct$ model, sometimes refereed to as a {\it linearly coasting cosmology}. A number of the objections to the $R_h = ct$ cosmology have been presented previously in terms of the analysis of power-law cosmologies \citep[e.g.][]{1989ApJ...344..543K,2008arXiv0804.3491D}, although these appear to be absent from the discussion in this latest incarnation of the linearly expanding cosmological model. 
Various objections to the predictions of power-law cosmologies were presented by \citet{1999PhRvD..59d3514K}, responding to claims in earlier papers, noting several observational results that are directly in conflict with the predictions from the $R_h = ct$ cosmological model. In the following, we focus upon the implications of linear expansion on nucleosynthesis, considering several points discussed in previous contributions.

\subsection{The Simmering Universe}\label{simmer}
While the discussion of primordial nucleosynthesis in an $R_h = ct$ cosmological model has not been presented in the literature, at an oral presentation it was suggested that the idea of a "simmering universe", gently cooking up elements over an extended period, might enable the production of sufficient quantities of helium in the linearly expanding universe (Melia, private communication to Barnes). 
This model was derived in the original discussion of power-law cosmologies, although explicit detail in the literature is somewhat lacking \citep{2000IJMPD...9..757B,2002PhLB..548...12D}. This acknowledges that weak interactions remain in thermal equilibrium for significantly longer period in power-law cosmologies than in standard cosmological models and that the neutron abundance is significantly depleted by the onset of nucleosynthesis. 
The claim is, however, once the production of $^\text{2}$H overcomes  photo-destruction, then inverse $\beta$-decay of protons will maintain the production of helium, bringing the resultant abundance to the observed level. A precise analysis of the approach adopted in modelling the `simmering universe' in previous works is difficult,as details are scant, although it is stated that numerical realisations were undertake with the  {\tt NUC123} fortran code \citep{1992STIN...9225163K} with the required integrations apparently required quadruple precision calculations. To achieve the required helium yield, the baryon-to-photon ratio, $\eta$, had to be increased to $\sim10^{-8}$, more than an order of magnitude larger than the measured value of $6.19\times10^{-10}$ \citep{2014A&A...571A..16P}.

\citet{1999PhRvD..59d3514K} and \citet{2000PhRvD..61j3507K}  also considered primordial nucleosynthesis in a linearly expanding cosmological model, partly in response to further support of the ``simmering universe'' by \citet{1999PhRvD..60j8301S}, again finding against this cosmological model by showing that the depletion of neutrons at nucleosynthesis results in a substantial deficit in the results abundance of helium. They also found that the helium mass fraction can be enhanced to be consistent with observational constraints by substantially boosting the baryon to photon ratio, although the resultant abundances of deuterium, $^\text{3}$He and $^\text{7}$Li are significantly depleted at $\sim10^{-18}$, $\sim10^{-12}$ and $\sim10^{-9}$ respectively.

\begin{figure}
		\includegraphics[width=0.95\columnwidth]{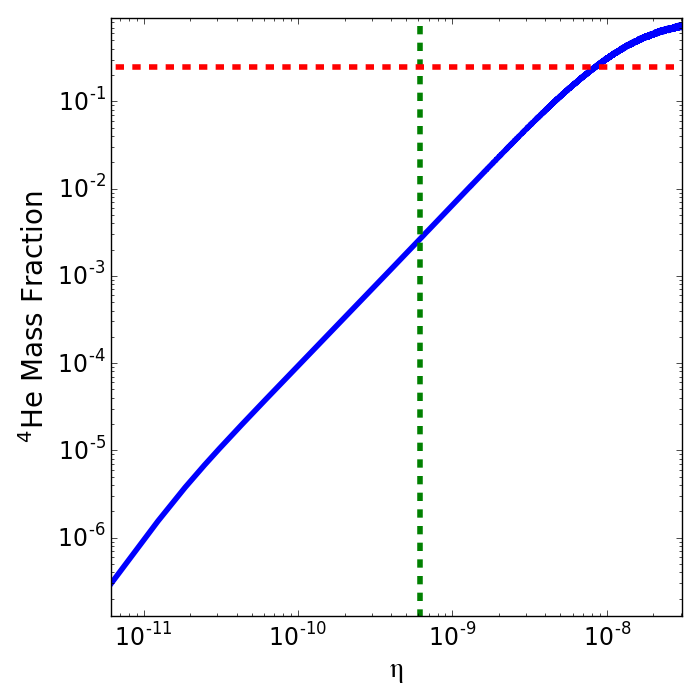}
    \caption{The mass fraction of ${\rm ^4He}$ as a function of the baryon to photon ratio (blue line). The green vertical line is the observed baryon to photon ratio, whereas the red horizontal line is the observed mass fraction.}
    \label{heliumyield}
\end{figure}

We repeated this analysis for a simmering universe using our modified version of {\tt AlterBBN}, maintaining the expansion history and temperature dependence outlined previously (Section~\ref{numerical}) but considering the differing values of $\eta$. The results of this appear in Figure~\ref{heliumyield}, which presents the $^\text{4}$He mass fraction as a function of $\eta$. The green dashed vertical line represents the observed value of $\eta = 6.19 \times 10^{-10}$, where as the red dashed horizontal line is the observed helium mass fraction of $0.2470$ \citep{2015arXiv150501076C}. As discussed above, at the observed value of $\eta$, the $R_h = ct$ cosmology results in a helium mass fraction of $\sim10^{-3}$, far below the observed value. Clearly, decreasing $\eta$ results in an even smaller helium mass factor, but increasing the baryon to photon ratio yields a higher resultant helium mass fraction. Inline with previous claims, $\eta$ has to be increased to around $\sim10^{-8}$, an order of magnitude above the observed limit, to achieve this substantial production of helium.

\begin{figure}
		\includegraphics[width=0.99\columnwidth]{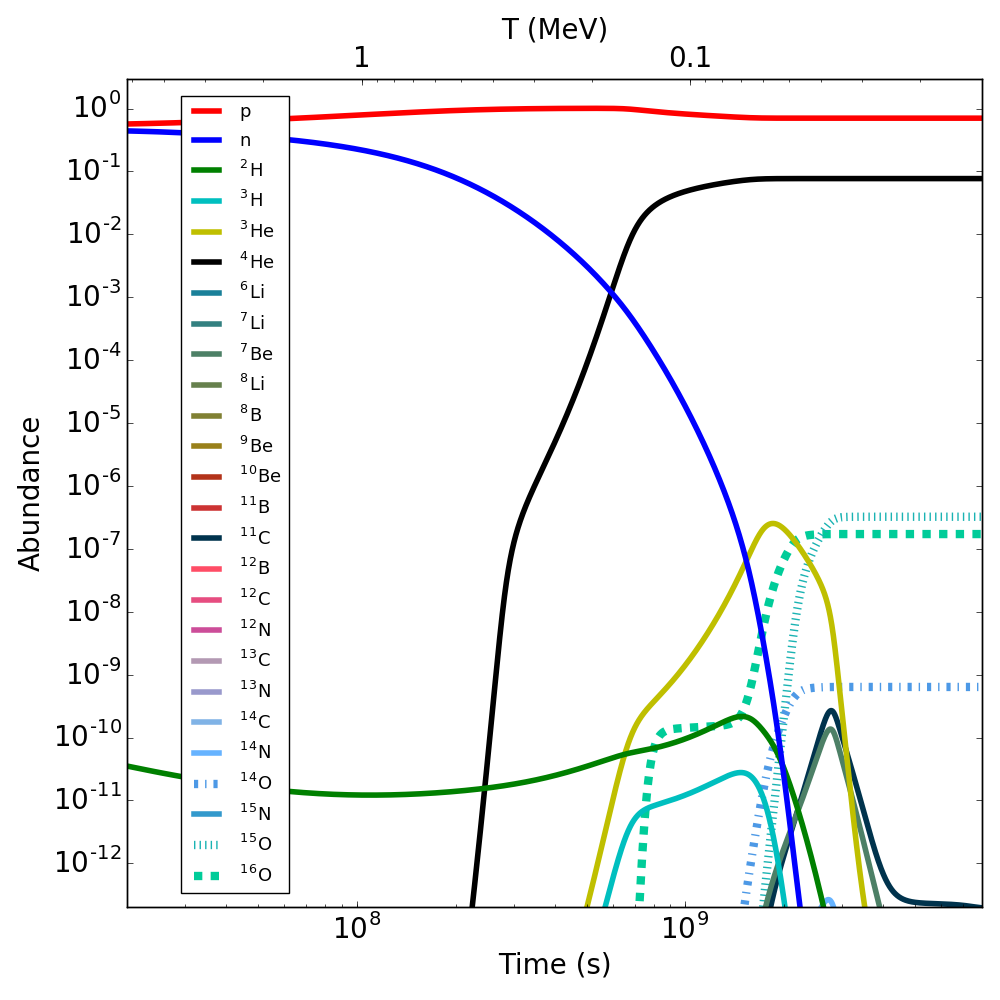}
    \caption{As for Figure~\ref{abundance_vs_time}, but now considering a baryon to photon ratio of $\eta\sim10^{-8}$. The entire element output of {\tt AlterBBN} is displayed. Note that while an appreciable quantity of 
    $^\text{4}$He is produced, other light elements are rapidly consumed and burnt into heavier elements, resulting in significant quantities of $^\text{14}$O, $^\text{15}$O and $^\text{16}$O. We caution the reader, however, that the 88 reactions considered in this code only include those up to these elements and hence we currently do not know if this is the true end point of nucleosynthesis in the simmering universe.}
    \label{abundance_vs_time_eta}
\end{figure}

Given the extended period over which the ``simmering universe'' has to produce helium, it is interesting to consider the abundances of other elements in such a universe. In the following, examine nucleosynthesis in $R_h = ct$ cosmology, considering an boosted baryon to photon ratio of $\eta\sim10^{-8}$, with the results being presented in Figure~\ref{abundance_vs_time_eta}. As noted above, with this, appreciable quantities are produced during nucleosynthesis, bringing the abundance into line with observations. However, this increase in the baryon to photon ratio significantly impacts the rest of element production in the early universe. Again, deuterium abundance rapidly decreases beyond $\sim2\times10^9$ seconds, in conflict with observations. However, there are other intriguing features, with many of the heavier elements diminishing after a few times $10^9$ seconds, leaving them absent from the universe. Clearly, in the long times of nucleosynthesis in this enhanced baryon to photon universe, these lighter elements are being burnt into heavier and heavier elements. As seen in Figure~\ref{abundance_vs_time_eta}, at the end of nucleosynthesis, the only appreciable elemental abundances are $^\text{1}$H and $^\text{4}$He, with smaller amounts of $^\text{14}$O, $^\text{15}$O and $^\text{16}$O. 

At this point, it is worth noting that this conclusion is somewhat different to that presented by \citet{2000PhRvD..61j3507K} who found that, while not substantial, their modelling suggested that the simmering $R_h = ct$ universe produces some light elements. To understand this, it is important to understand that their considered following the abundances of light elements in statistical equilibrium at high temperatures until they reach freeze-out values. However, they considered a limited number of elements and rates in their analysis, limiting nucleosynthesis to the production of the lightest elements.
The {\tt AlterBBN} algorithm, on the other hand, considers a total of 88 reaction rates to calculate the evolution of the elemental abundances through nucleosynthesis \citep{2012CoPhC.183.1822A}, and hence provides pathways to heavier elements that were not considered in this earlier analysis, something simply revealed by "switching-off" some of the reactions in our numerical approach. Hence, if \citet{2000PhRvD..61j3507K}  had considered these additional reactions, they too would have found that the deuterium, $^\text{3}$He and $^\text{7}$Li they found after nucleosynthesis were burnt to heavier elements as seen in Figure~\ref{abundance_vs_time_eta}. It is important, however, to insert a word of caution into the element abundances found after nucleosynthesis in this present study. In the 88 reactions considered in {\it AlterBBN}, the heaviest elements produced are the isotopes of oxygen. Hence, it is quite likely that final oxygen abundances present in Figure~\ref{abundance_vs_time_eta} are due to the lack of inclusion of other reaction pathways to yet heavier elements, and we should expect that this simmering universe, very heavy elements are produced. Including these reaction paths into {\tt AlterBBN} is beyond the scope of this work, but clearly this burning of all light elements, other than hydrogen and helium, into heavier elements is catastrophic for this cosmology and  this is not a route to save the $R_h = ct$ cosmological model. 

\section{Conclusions}\label{conclusions}
We have considered primordial nucleosynthesis within the recently proposed $R_h = ct$ universe, a cosmology that demands linear evolution of the scale-factor at all times. We show that, compared to the standard $\Lambda$CDM cosmology, this universe cools extremely slowly, with the period of nucleosynthesis occurring several months to years after the Big Bang, rather than the seconds to minutes we encounter typical cosmological models. The long-term cooling has a significant influence on the state of the universe just before nucleosynthesis, as the $p \leftrightarrow n$ reaction rates remaining in equilibrium over a substantial time, meaning that there is no freeze-out of the proton-to-neutron abundance. The implication of this is that at the onset of nucleosynthesis in ernest the proton-to-neutron abundance is $\sim 10^{-2} - 10^{-3}$, rather than the $\sim\frac{1}{6}$ in standard cosmological models. This lack of available neutrons feeds into the abundance of deuterium, and then into the production of ${\rm ^4He}$, with a resulting abundance of $\sim10^{-3}$, differing significantly from the observed mass fraction of  $\sim$25\%. Furthermore, the other resulting elemental abundances are also in conflict with the observed values.

We showed that the `simmering universe' can result in an appreciable helium mass fraction if the baryon to photon ratio, $\eta$ is boosted to $\sim10^{-8}$, more than order of magnitude larger than the observed value. However, the sustained period of nucleosynthesis in this model results in the synthesis of substantial quantities of heavier elements, robbing the universe of any lighter elements. The limitations of the nuclear pathways considered in this study have shown this up to oxygen and it is envisaged that this continues to heavier elements. Hence, the $R_h = ct$ cosmological model appears to be incapable of producing a realistic primordial abundance of elements.

In closing, we note that one of the apparent strngths of the $R_h = ct$ cosmological model is some how "simpler" than the standard cosmological model \citep{2012arXiv1205.2713M}, although this simplicity hides the completely unknown internal physics of the dark sector to demand a linear expansion \citep{2013MNRAS.432.2324L,2015MNRAS.446.1191M}. The inability of this model to reproduce anywhere near the required abundances of elements in primordial nucleosynthesis is also very troubling, and further additional physics is  needed in the early universe to attempt to bring this cosmology into agreement with observations. Whatever the case, this  ``simple'' cosmological model does not appear to be so simple after all.

\section*{Acknowledgements}
We would like to thank A. Arbey for discussions on his {\tt AlterBBN} software and for making it public. K. Bolejko is thanked for interesting discussions and comments. The initial stages of this research were undertaken by R. Kaushik as part of his honours project at the University of Sydney.





\begin{thebibliography}{99}
%
\bibitem[\protect\citeauthoryear{Planck Collaboration et al.}{2014}]{2014A&A...571A..16P} Planck Collaboration, et al., 2014, A\&A, 571, A16 
%
\bibitem[\protect\citeauthoryear{Alpher, Herman, 
\& Gamow}{1948}]{1948PhRv...74.1198A} Alpher R.~A., Herman R., Gamow G.~A., 1948, PhRv, 74, 1198 
%
\bibitem[\protect\citeauthoryear{Arbey}{2012}]{2012CoPhC.183.1822A} Arbey 
A., 2012, CoPhC, 183, 1822 
%
\bibitem[\protect\citeauthoryear{Batra et al.}{2000}]{2000IJMPD...9..757B} 
Batra A., Lohiya D., Mahajan S., Mukherjee A., Ashtekar A., 2000, IJMPD, 9, 757 
%
\bibitem[\protect\citeauthoryear{Bilicki \& Seikel}{2012}]{2012MNRAS.425.1664B} Bilicki M., Seikel M., 2012, MNRAS, 425, 1664 
%
\bibitem[\protect\citeauthoryear{Bikwa, Melia, \& Shevchuk}{2012}]{2012MNRAS.421.3356B} Bikwa O., Melia F., Shevchuk A., 2012, MNRAS, 421, 3356 
%
\bibitem[\protect\citeauthoryear{Cooke et al.}{2014}]{2014ApJ...781...31C} Cooke R.~J., Pettini M., Jorgenson R.~A., Murphy M.~T., Steidel C.~C., 2014, ApJ, 781, 31
%
\bibitem[\protect\citeauthoryear{Cyburt et al.}{2015}]{2015arXiv150501076C} Cyburt R.~H., Fields B.~D., Olive K.~A., Yeh T.-H., 2015, arXiv, arXiv:1505.01076
%
\bibitem[\protect\citeauthoryear{Dev, Jain, 
\& Lohiya}{2008}]{2008arXiv0804.3491D} Dev A., Jain D., Lohiya D., 2008, arXiv, arXiv:0804.3491
%
\bibitem[\protect\citeauthoryear{Dev et al.}{2002}]{2002PhLB..548...12D} 
Dev A., Safonova M., Jain D., Lohiya D., 2002, PhLB, 548, 12 
%
\bibitem[\protect\citeauthoryear{Harrison}{1991}]{1991ApJ...383...60H} Harrison E., 1991, ApJ, 383, 60 
%
\bibitem[Hoyle 
\& Tayler(1964)]{1964Natur.203.1108H} Hoyle, F., \& Tayler, R.~J.\ 1964, \nat, 203, 1108 
%
\bibitem[\protect\citeauthoryear{Kaplinghat et 
al.}{1999}]{1999PhRvD..59d3514K} Kaplinghat M., Steigman G., Tkachev I., 
Walker T.~P., 1999, PhRvD, 59, 043514 
%
\bibitem[\protect\citeauthoryear{Kaplinghat, Steigman, \& Walker}{2000}]{2000PhRvD..61j3507K} Kaplinghat M., Steigman G., Walker T.~P., 2000, PhRvD, 61, 103507 
%
\bibitem[\protect\citeauthoryear{Kawano}{1992}]{1992STIN...9225163K} Kawano 
L., 1992, STIN, 92
%
\bibitem[\protect\citeauthoryear{Kim, Lasenby, \& Hobson}{2016}]{2016arXiv160107890K} Kim D.~Y., Lasenby A.~N., Hobson M.~P., 2016, arXiv, arXiv:1601.07890 
%
\bibitem[\protect\citeauthoryear{Kolb}{1989}]{1989ApJ...344..543K} Kolb 
E.~W., 1989, ApJ, 344, 543
%
\bibitem[\protect\citeauthoryear{Kolb 
\& Turner}{1990}]{1990eaun.book.....K} Kolb E.~W., Turner M.~S., 1990, "The Early Universe", Westview Press
%
\bibitem[\protect\citeauthoryear{Kumar 
\& Lohiya}{2008}]{2008arXiv0802.1124K} Kumar P., Lohiya D., 2008, arXiv, arXiv:0802.1124 
%
\bibitem[\protect\citeauthoryear{Lewis 
\& van Oirschot}{2012}]{2012MNRAS.423L..26L} Lewis G.~F., van Oirschot P., 2012, MNRAS, 423, L26 
%
\bibitem[\protect\citeauthoryear{Lewis}{2013}]{2013MNRAS.431L..25L} Lewis 
G.~F., 2013, MNRAS, 431, L25 
%
\bibitem[\protect\citeauthoryear{Lewis}{2013}]{2013MNRAS.432.2324L} Lewis 
G.~F., 2013, MNRAS, 432, 2324
%
\bibitem[\protect\citeauthoryear{Melia}{2007}]{2007MNRAS.382.1917M} Melia F., 2007, MNRAS, 382, 1917 
%
\bibitem[\protect\citeauthoryear{Melia \& Abdelqader}{2009}]{2009IJMPD..18.1889M} Melia F., Abdelqader M., 2009, IJMPD, 18, 1889 
%
\bibitem[\protect\citeauthoryear{Melia}{2012a}]{2012AJ....144..110M} Melia F., 2012a, AJ, 144, 110 
%
\bibitem[Melia(2012b)]{2012arXiv1205.2713M} Melia, F.\ 2012b, Australian Journal of Physics (May) 
%
\bibitem[\protect\citeauthoryear{Melia \& Shevchuk}{2012}]{2012MNRAS.419.2579M} Melia F., Shevchuk A.~S.~H., 2012, MNRAS, 419, 2579 
%
\bibitem[\protect\citeauthoryear{Melia}{2013}]{2013A&A...553A..76M} Melia F., 2013, A\&A, 553, A76 
%
\bibitem[\protect\citeauthoryear{Melia}{2015}]{2015MNRAS.446.1191M} Melia F., 2015, MNRAS, 446, 1191 
%
\bibitem[\protect\citeauthoryear{Melia \& McClintock}{2015}]{2015AJ....150..119M} Melia F., McClintock T.~M., 2015, AJ, 150, 119 
%
\bibitem[\protect\citeauthoryear{Melia \& Fatuzzo}{2016a}]{2016MNRAS.456.3422M} Melia F., Fatuzzo M., 2016a, MNRAS, 456, 3422 
%
\bibitem[\protect\citeauthoryear{Melia}{2016b}]{2016FrPhy..11..557M} Melia F., 2016b, FrPhy, 11, \#557 
%
\bibitem[\protect\citeauthoryear{Melia}{2016c}]{2016arXiv160201435M} Melia F., 2016c, arXiv, arXiv:1602.01435 
%










\bibitem[\protect\citeauthoryear{Mitra}{2014}]{2014MNRAS.442..382M} Mitra A., 2014, MNRAS, 442, 382 
%
\bibitem[Mukhanov(2005)]{2005pfc..book.....M} Mukhanov, V.\ 2005, Physical 
Foundations of Cosmology, by Viatcheslav Mukhanov, pp.~442.~Cambridge 
University Press, November 2005.~ISBN-10: 0521563984.~ISBN-13: 
9780521563987.~LCCN: QB981 .M89 2005, 442 
%
\bibitem[Padmanabhan(2006)]{2006AIPC..843..111P} Padmanabhan, T.\ 2006, Graduate School in Astronomy: X, 843, 111 
%
\bibitem[\protect\citeauthoryear{Peebles}{1966}]{1966ApJ...146..542P} Peebles P.~J.~E., 1966, ApJ, 146, 542 
%
\bibitem[\protect\citeauthoryear{Shafer}{2015}]{2015PhRvD..91j3516S} Shafer D.~L., 2015, PhRvD, 91, 103516 
%
\bibitem[\protect\citeauthoryear{Sethi, Batra, \& Lohiya}{1999}]{1999PhRvD..60j8301S} Sethi M., Batra A., Lohiya D., 1999, PhRvD, 60, 108301 
%
\bibitem[\protect\citeauthoryear{Smith, Kawano, 
\& Malaney}{1993}]{1993ApJS...85..219S} Smith M.~S., Kawano L.~H., Malaney R.~A., 1993, ApJS, 85, 219 
%
\bibitem[\protect\citeauthoryear{Smith 
\& Fuller}{2010}]{2010PhRvD..81f5027S} Smith C.~J., Fuller G.~M., 2010, PhRvD, 81, 065027 
%
\bibitem[\protect\citeauthoryear{van Oirschot, Kwan, 
\& Lewis}{2010}]{2010MNRAS.404.1633V} van Oirschot P., Kwan J., Lewis G.~F., 2010, MNRAS, 404, 1633 
%
\bibitem[van Oirschot et al.(2015)]{2015mgm..conf.1567V} van Oirschot, P., 
Kwan, J., 
\& Lewis, G.~F.\ 2015, Thirteenth Marcel Grossmann Meeting: On Recent Developments in Theoretical and Experimental General Relativity, Astrophysics and Relativistic Field Theories, 1567 
%
\bibitem[\protect\citeauthoryear{Wagoner, Fowler, 
\& Hoyle}{1967}]{1967ApJ...148....3W} Wagoner R.~V., Fowler W.~A., Hoyle F., 1967, ApJ, 148, 3 
%
\bibitem[\protect\citeauthoryear{Wagoner}{1968}]{1968ApJ...151L.103W} 
Wagoner R.~V., 1968, ApJ, 151, L103 
%
\end{thebibliography}


\bsp	
\label{lastpage}
\end{document}